\documentclass[12pt,twoside]{article}
\usepackage{a4}

\begin{document}

\newcommand{\gf}{G_{\mbox{{\scriptsize F}}}}
\newcommand{\order}{{\cal O}}
\newcommand{\delp}{\Delta P}
\newcommand{\leqsim}{\,\mbox{{\scriptsize $\stackrel{<}{\sim}$}}\,}
\newcommand{\geqsim}{\,\mbox{{\scriptsize $\stackrel{>}{\sim}$}}\,}
\newcommand{\bskk}{B_s\to K^+ K^-}
\newcommand{\bsokk}{B_s^0\to K^+ K^-}
\newcommand{\bsobkk}{\bar B_s^0\to K^+ K^-}
\newcommand{\bppipko}{B^+\to \pi^+ K^0}
\newcommand{\bmpimkob}{B^-\to \pi^- \bar K^0}
\newcommand{\bmpiokm}{B^-\to \pi^0 K^-}
\newcommand{\bppiokp}{B^+\to \pi^0 K^+}
\newcommand{\bdpimkp}{B^0_d\to \pi^- K^+}
\newcommand{\bdbpipkm}{\bar B^0_d\to \pi^+ K^-}
\newcommand{\pew}{P_{\mbox{{\scriptsize EW}}}}
\newcommand{\pewb}{\bar P_{\mbox{{\scriptsize EW}}}}
\newcommand{\pewp}{P_{\mbox{{\scriptsize EW}}}'}
\newcommand{\pewc}{P_{\mbox{{\scriptsize EW}}}^{\mbox{{\scriptsize C}}}}
\newcommand{\pcew}{P_{\mbox{{\scriptsize EW}}}^{\mbox{{\scriptsize (C)}}}}
\newcommand{\pcewb}{\bar P_{\mbox{{\scriptsize
EW}}}^{\mbox{{\scriptsize (C)}}}}
\newcommand{\pewpc}{P_{\mbox{{\scriptsize EW}}}'^{\mbox{{\scriptsize C}}}}
\newcommand{\bbtodb}{\bar b\to\bar d}
\newcommand{\bbtosb}{\bar b\to\bar s}

\newcommand{\bkk}{B_d\to K^0\bar K^0}
\newcommand{\bksks}{B_d\to K_{\mbox{{\scriptsize S}}}
K_{\mbox{{\scriptsize S}}}}
\newcommand{\bpiphi}{B_s\to\pi^0\Phi}
\newcommand{\bopiphi}{B_s^0\to\pi^0\Phi}
\newcommand{\bbopiphi}{\bar B_s^0\to\pi^0\Phi}
\newcommand{\brhok}{B_s\to\rho^0K_{\mbox{{\scriptsize S}}}}
\newcommand{\bqf}{B_q\to f}
\newcommand{\pcps}{\phi_{\mbox{{\scriptsize CP}}}(B_s)}
\newcommand{\pcpq}{\phi_{\mbox{{\scriptsize CP}}}(B_q)}
\newcommand{\pw}{\phi_{\mbox{{\scriptsize W}}}}
\newcommand{\acp}{a_{\mbox{{\scriptsize CP}}}}
\newcommand{\acpdir}{{\cal A}_{\mbox{{\scriptsize CP}}}^
{\mbox{{\scriptsize dir}}}}
\newcommand{\acpmi}{{\cal A}_{\mbox{{\scriptsize
CP}}}^{\mbox{{\scriptsize mix-ind}}}}
\newcommand{\acc}{A_{\mbox{{\scriptsize CC}}}}
\newcommand{\aew}{A_{\mbox{{\scriptsize EWP}}}}
\newcommand{\heff}{{\cal H}_{\mbox{{\scriptsize eff}}}(\Delta B=-1)}
\newcommand{\xif}{\xi_f^{(q)}}
\newcommand{\xiskk}{\xi_{K^+K^-}^{(s)}}
\newcommand{\xikk}{\xi_{K^0\bar K^0}^{(d)}}
\newcommand{\VmA}{\mbox{{\scriptsize V--A}}}
\newcommand{\VpA}{\mbox{{\scriptsize V+A}}}
\newcommand{\VpmA}{\mbox{{\scriptsize V$\pm$A}}}
\newcommand{\beq}{\begin{equation}}
\newcommand{\eeq}{\end{equation}}
\newcommand{\bea}{\begin{eqnarray}}
\newcommand{\eea}{\end{eqnarray}}
\newcommand{\non}{\nonumber}
\newcommand{\lab}{\label}
\newcommand{\la}{\langle}
\newcommand{\ra}{\rangle}
\newcommand{\np}{Nucl.\ Phys.}
\newcommand{\pl}{Phys.\ Lett.}
\newcommand{\prl}{Phys.\ Rev.\ Lett.}
\newcommand{\pr}{Phys.\ Rev.}
\newcommand{\zp}{Z.\ Phys.}

\setcounter{page}{-1}
\thispagestyle{empty}
\begin{flushright}
TUM-T31-95/95\\
MPI-PhT/95-68\\
TTP95-29\\
hep-ph/9507303\\
July 1995
\end{flushright}

\begin{center}
\vspace{0.2cm}
{\Large{\bf Towards the Control over Electroweak Penguins}}\\
\vspace{0.5cm}
{\Large{\bf in Nonleptonic $B$-Decays
\footnote[1]{Supported in part by the German {\it Bundesministerium
f\"ur Bildung und Forschung} under contract 06--TM--743 and
by the CEC science project SC1--CT91--0729.}}}\\
\vspace{0.9cm}
{\large{\sc Andrzej J. Buras}}\\

\vspace{0.1cm}
{\sl Technische Universit\"at M\"unchen, Physik Department\\
D--85748 Garching, Germany}\\
\vspace{0.3cm}
{\sl Max-Planck-Institut f\"ur Physik\\
-- Werner-Heisenberg-Institut --\\
F\"ohringer Ring 6, D--80805 M\"unchen, Germany}\\

\vspace{0.9cm}

{\large{\sc Robert Fleischer}}\\
\vspace{0.1cm}
{\sl Institut f\"ur Theoretische Teilchenphysik\\
Universit\"at Karlsruhe\\
D--76128 Karlsruhe, Germany}\\
\vspace{0.9cm}
{\large{\bf Abstract}}
\end{center}
\vspace{0.3cm}
We present strategies for determining electroweak penguins from
experimental data. Using the CKM-angle $\gamma$ as one of our central
inputs and making some reasonable approximations, we show that the
$\bar b\to\bar s$ electroweak penguin amplitude can be determined in a
two-step procedure involving i) BR$(B^+\to\pi^0K^+)$, BR$(B^-\to\pi^0K^-)$,
BR$(B^+\to\pi^+K^0)$ and ii) either BR$(B^0_d\to\pi^-K^+)$,
BR$(\bar B^0_d\to\pi^+K^-)$ or $\acp(t)$ of the mode $B_s\to K^+K^-$.
The determination employing the $B\to\pi K$ transitions is not affected
by $SU(3)$-breaking effects. Relating the $\bar b\to\bar s$ electroweak
penguin amplitude to the $\bar b\to\bar d$ case through $SU(3)$ symmetry
arguments, we are in a position to estimate the electroweak penguin
uncertainty affecting the extraction of the CKM-angle $\alpha$ by using
isospin relations among $B\to\pi\pi$ decays. Our results allow in
principle the determination of CKM-phases in a variety of $B$-decays.

\newpage
\thispagestyle{empty}
\mbox{}
\newpage

During the last two years there has been a considerable interest in
the role of electroweak penguin contributions in non-leptonic $B$-decays.
Since the Wilson coefficients of the corresponding local operators
increase strongly with the top-quark mass, it has been found
\cite{rfewp1,rfewp2,rfewp3} that the role of the electroweak
penguins can be substantial in certain decays. This is for instance
the case of the decay $B^-\to K^-\Phi$ \cite{rfewp1},
which  exhibits sizable electroweak penguin effects.
More interestingly, there are even some channels, such as $B^-\to\pi^-\Phi$
\cite{rfewp2} and $B_s\to\pi^0\Phi$ \cite{rfewp3}, which are
{\it dominated completely} by electroweak penguin contributions
and which should, thus, allow interesting insights into the physics
of the corresponding operators. In this respect, the
decay $B_s\to\pi^0\Phi$ (or similar transitions such as $B_s\to\rho^0\Phi$)
is very promising due to its special isospin-, CKM- and
colour-structure \cite{rfewp3}. As the branching ratio of this mode
is expected to be of $\order(10^{-7})$, it will unfortunately be
rather difficult to analyze this decay experimentally.
The electroweak penguin effects discussed in refs.~\cite{rfewp1,rfewp3}
have been confirmed by other authors \cite{dhewp1}-\cite{dy}.

In the foreseeable future the branching ratios of ${\cal O}(10^{-5})$
and possibly ${\cal O}(10^{-6})$ will be experimentally available and
it is important to ask about the role of electroweak penguin effects
in the corresponding channels. In particular, the question arises whether
the usual strategies for the determination of the CKM-phases are
affected by the presence of the electroweak penguin contributions.

It is evident that the pure tree diagram decays do not receive any
contributions from electroweak penguins. Consequently,
the very clean method for the determination of the phase $\gamma$
proposed by Gronau and Wyler \cite{gw} involving charged $B$-decays
of the type $B^\pm\to DK^\pm$ (see also ref.~\cite{dun}) remains
unaffected by these new contributions. This applies also to the
$\gamma$-determination
proposed by Aleksan et al.~\cite{adk} which uses a measurement of the
time-dependent decay rates of the transitions $B_s\to D_s^\pm K^\mp$.
Similar comments apply to the ``gold-plated'' decay $B_d\to
\psi K_{\mbox{{\scriptsize S}}}$ in which the electroweak penguins
having the same phase as the leading
tree contribution do not obscure a very clean determination of the
phase $\beta$.

The situation concerning the $\alpha$-determination by means of the
isospin relations among $B\to\pi\pi$ decays proposed by Gronau
and London~\cite{gl}
is more involved, however. As pointed out first by Deshpande and He
\cite{dhewp2}, the impact of electroweak penguins on this determination
could be sizable. A closer look \cite{ghlrewp} shows, however, that
this impact is rather small, at most a few $\%$. On the other hand, it is
now well accepted \cite{dhewp2, ghlrewp} that the electroweak penguins
should have a considerable impact on the  methods proposed last year by
Gronau, Hern\'andez, London and Rosner \cite{grl}-\cite{ghlrsu3}
to measure both weak and strong phases
by using $SU(3)$ triangle relations among $B\to\{\pi\pi,\pi
K,K\bar K\}$ decays and making certain plausible dynamical assumptions
(e.g.\ neglect of annihilation topologies).

While this point has been shown explicitly in ref.~\cite{dhewp2}, a
systematic classification of electroweak penguins in two-body
$B$-decays has been presented in ref.~\cite{ghlrewp}. Moreover,
in this paper, Gronau et al.\ have constructed an amplitude quadrangle
for $B\to\pi K$ decays that can be used -- at least in principle --
to extract the CKM-angle $\gamma$ irrespectively of the presence of
electroweak penguins. Unfortunately, from the experimental point of view
this approach is rather difficult, because one \mbox{diagonal} of the
quadrangle corresponds to the amplitude of the electroweak penguin
dominated $B_s$-decay $B_s\to\pi^0\eta$ which is expected to have
a very small branching ratio at the $\order(10^{-7})$ level. Another
$SU(3)$-symmetry based method of extracting $\gamma$, where electroweak
penguins are also eliminated, has been presented very recently by
Deshpande and He~\cite{dhgam}. Although this approach using the
charged $B$-decays $B^-\to\{\pi^-\bar K^0,\pi^0K^-,\eta K^-\}$ and
$B^-\to\pi^-\pi^0$ should be more promising for experimentalists,
it is affected by $\eta$--$\eta'$--mixing and other $SU(3)$-breaking
effects and therefore cannot be regarded as a clean measurement
of $\gamma$.

In view of this situation, it would be useful
to determine the electroweak penguin contributions experimentally.
Once this has been achieved, their role in a variety of $B$-decays
could be explicitly found. Although some thoughts on this issue have
appeared in~\cite{ghlrewp}, no constructive quantitative
method has been proposed there.

Here we would like to suggest a
different ``philosophy'' of applying the $SU(3)$ amplitude relations.
In contrast to Gronau et al., we think that these relations are
more useful from the phenomenological point of view if one uses
the phase $\gamma$ {\it as one of the central inputs}.
As we have stated above,
there are already methods on the market allowing a measurement
of this phase in an {\it absolutely clean way} without any effect
coming from the electroweak penguins.
Although these methods (for a review see e.g.\ ref.~\cite{kay})
are quite difficult from the experimental point of view as well, they
should be easier for experimentalists than the quadrangle of
ref.~\cite{ghlrewp}.

At first sight, this new philosophy might appear not useful because
one of the goals of the GHLR strategy was precisely the
determination of $\gamma$. Yet, as we have seen, this program
is difficult to realize without further inputs. On the other
hand, as we will show below, once the phase $\gamma$ is used as an
input, the electroweak penguin contributions can be straightforwardly
determined. This knowledge subsequently allows the determination
of CKM-phases in a variety of $B$-decays \cite{bfewpapp}. Consequently,
with this new strategy, the GHLR method is resurrected. Moreover, the
impact of electroweak penguins on the $\alpha$-determination using
$B(\bar B)\to\pi\pi$ decays can be \mbox{{\it quantitatively}} estimated.

The central point of this letter is a strategy for  determining the
$\bbtosb$ electroweak penguin amplitude from experimental data.
Whereas electroweak $\bbtodb$ penguins are expected to be rather small
in the case of $B$-decays into two-pion final states, the corresponding
$\bbtosb$ electroweak penguins are expected to affect $B\to\pi K$
transitions significantly \cite{dhewp2,ghlrewp}. Besides the knowledge
of the CKM-angle $\gamma$ our approach involves certain approximations
that will be discussed in a moment.

Let us begin our analysis by considering the $B$-meson decays
$\bppipko$, $\bppiokp$, $\bdpimkp$ and, moreover, the $B_s$-transition
$\bskk$. Applying the $SU(3)$ flavour symmetry of strong interactions
and using the same notation as Gronau, Hern\'andez, London and Rosner
in ref.~\cite{ghlrewp}, the corresponding decay amplitudes take the form
\beq\lab{e1}
\begin{array}{rcl}
A(\bppipko)&=&P'+c_d\pewpc\\
A(\bppiokp)&=&-\frac{1}{\sqrt{2}}\left[P'+T'+(c_u-c_d)\pewp+C'+c_u\pewpc
\right]\\
A(\bdpimkp)&=&-(P'+T'+c_u\pewpc)\\
A(\bsokk)&=&-(P'+T'+c_u\pewpc),
\end{array}
\eeq
where $T'$ and $C'$ describe colour-allowed and colour-suppressed
$\bar b\to\bar uu\bar s$ tree-level amplitudes, respectively, $P'$
denotes $\bbtosb$ QCD penguins, $\pewp$ is related to colour-allowed
$\bbtosb$ electroweak penguins and $\pewpc$ to colour-suppressed
electroweak penguins. Following the plausible arguments of Gronau
et al.\ outlined in refs.~\mbox{\cite{ghlrewp,ghlrsu3}}, we expect the
following hierarchy of the different topologies given in eq.~(\ref{e1}):
\beq\lab{e2}
\begin{array}{rcl}
1&:&|P'|\\
\order(\bar\lambda)&:&|T'|, \quad \left|\pewp\right|\\
\order(\bar\lambda^2)&:&|C'|, \quad \left|\pewpc\right|.
\end{array}
\eeq
Note that the parameter $\bar\lambda=\order(0.2)$ appearing in these
relations is not related to the usual Wolfenstein parameter $\lambda$.
It has been introduced by Gronau et al.\ just to keep track of the
expected orders of magnitudes. In eq.~(\ref{e2}), we have named this
quantity $\bar\lambda$ in order not to confuse it with Wolfenstein's
$\lambda$.

Consequently, if we neglect the colour-suppressed electroweak penguin
contributions $\pewpc$, which will simplify our
analysis considerably, the $C'$ amplitudes have to be neglected as
well since both topologies are expected to be of the same order in
$\bar\lambda$. Within this approximation, we obtain
\beq\lab{e3}
\begin{array}{rcl}
A(\bppipko)&=&P'\\
A(\bppiokp)&=&-\frac{1}{\sqrt{2}}\left[P'+T'+(c_u-c_d)\pewp\right]\\
A(\bdpimkp)&=&-(P'+T')\\
A(\bsokk)&=&-(P'+T').
\end{array}
\eeq
Note that exchange and annihilation-type topologies, which have not
been written explicitly in eq.~(\ref{e1}), have also to be neglected
within this approximation since they are expected to be
$\leqsim\order(\bar\lambda^2)$ \cite{ghlrewp,ghlrsu3}.

Due to the special CKM-structure of the $\bbtosb$ penguins, we have
\cite{bf}
\beq\lab{e4}
\begin{array}{rcl}
P'&=&|P'|e^{i\delta_{P'}}e^{i\pi}=\bar P'\\
\pewp&=&\left|\pewp\right|e^{i\delta_{EWP'}}e^{i\pi}=\bar\pewp,
\end{array}
\eeq
where the phases $\delta$ are CP-conserving strong final state
interaction phases and $\pi$ represents the CP-violating weak
phase.

Let us next rescale the transition amplitudes of the decays
$\bppipko$ and $\bppiokp$ by a factor $|P'|$. Taking furthermore
into account the relation
\beq\lab{e5}
\bar T'=e^{-2i\gamma}T',
\eeq
one can easily draw Fig.~1 representing the first two decay amplitudes
given in eq.~(\ref{e3}) and those of the corresponding
CP-conjugate modes. Looking at this figure implies that the $\bbtosb$
electroweak penguin amplitude $(c_u-c_d)\pewp$ can be constructed by
measuring the rates of the decays $\bppiokp$, $B^-\to\pi^0K^-$ and
$\bppipko$, provided both the amplitude
\beq\lab{e6}
z\equiv\frac{T'}{|P'|}
\eeq
and the CKM-angle $\gamma$ are known. Note that the quantity $z$ is
given in the $x'$--$y'$--frame defined in Fig.~1 by the expression
\beq\lab{e7}
z=e^{-i\omega}\frac{|T'|}{|P'|}.
\eeq
The phase $\delta_{P'}$ determining the orientation of this frame
cannot be fixed. However, concerning our phenomenological
applications this quantity is irrelevant.

In the following discussion we shall present two different approaches
of determining $z$ making use of the decays $\bdpimkp$
($\bdbpipkm$) and $\bskk$, respectively. Let us describe the method
involving the $B_d$-modes first. Taking into account both eqs.~(\ref{e4})
and (\ref{e5}) and the expression for the amplitude $A(\bdpimkp)$
given in eq.~(\ref{e3}), we can easily construct the two
triangles shown in Fig.~2. As can be seen from this figure,
if the CKM-angle $\gamma$ is known, the amplitude $z=T'/|P'|$ can be
determined by measuring the rates of the decays $\bdpimkp$, $\bdbpipkm$
and $\bppipko$ which fixes $|P'|$. Note that this method requires no
time-dependent measurements and that all involved branching ratios
should be of $\order(10^{-5})$.

Let us now describe another independent approach of determining this
quantity which is more formal and requires a measurement of the
time-dependent CP-violating asymmetry of the mode $\bskk$. Since
this transition is the decay of a neutral $B_s$-meson into a
CP-eigenstate, the corresponding CP asymmetry is given by
\bea
\lefteqn{\acp(t)\equiv\frac{\Gamma(B_s^0(t)\to K^+K^-)-\Gamma(\bar
B_s^0(t)\to K^+K^-)}{\Gamma(B_s^0(t)\to K^+K^-)+\Gamma(\bar
B_s^0(t)\to K^+K^-)}=}\lab{e8}\\
&&\acpdir(\bskk)\cos(\Delta M_s
t)+\acpmi(\bskk)\sin(\Delta M_s t),\nonumber
\eea
where we have separated the {\it direct} CP-violating contributions,
which are proportional to
\beq\lab{e9}
\acpdir(\bskk)\equiv\frac{1-\left\vert\xiskk\right\vert^2}
{1+\left\vert\xiskk\right\vert^2},
\eeq
from those describing {\it mixing-induced} CP violation which are
characterized by
\beq\lab{e10}
\acpmi(\bskk)\equiv\frac{2\mbox{Im}\xiskk}{1+
\left\vert\xiskk\right\vert^2}.
\eeq
In eq.~(\ref{e8}), $\Delta M_s$ denotes the mass splitting of the
physical $B^0_s$--$\bar B^0_s$--mixing eigenstates.
The quantity $\xiskk$ containing essentially all the information
needed to evaluate the asymmetries (\ref{e9}) and (\ref{e10})
is given by
\beq\lab{e11}
\xiskk=-e^{-i0}\frac{A(\bsobkk)}{A(\bsokk)},
\eeq
where the factor $-e^{-i0}$ is related to $B^0_s$--$\bar B^0_s$--mixing.
Using eqs.~(\ref{e3}), (\ref{e4}) and writing the colour-allowed
tree-amplitude $T'$ in the form
\beq\lab{e12}
T'=|T'|e^{i\delta_{T'}}e^{i\gamma},
\eeq
where $\delta_{T'}$ is a strong phase shift and $\gamma$ is the usual
CKM-angle, we obtain
\bea
A(\bsokk)&=&-|P'|e^{i\delta_{P'}}\left[e^{i\pi}+\frac{|T'|}{|P'|}
e^{-i\omega}\right]\nonumber\\
A(\bsobkk)&=&-|P'|e^{i\delta_{P'}}\left[e^{i\pi}+\left(\frac{|T'|}{|P'|}
e^{-i\omega}\right)e^{-2i\gamma}\right],\lab{e13}
\eea
where $\omega$ is given by
\beq\lab{e14}
\omega=\delta_{P'}-\delta_{T'}-\gamma.
\eeq
Consequently, the quantity $|T'|/|P'|e^{-i\omega}$, which describes
the amplitude $z=T'/|P'|$ in the $x'$--$y'$--frame specified in
Fig.~1, is related to $\xiskk$ through the expression
\beq\lab{e15}
\frac{|T'|}{|P'|}e^{-i\omega}=\frac{1+\xiskk}{e^{-2i\gamma}+\xiskk}.
\eeq
If one measures the time-dependent CP asymmetry of the decay
$\bskk$, which is probably a rather difficult task for experimentalists
due to the large $B^0_s$--$\bar B^0_s$--mixing parameter
$x_s\equiv\tau_{B_s}\Delta M_s\geqsim10$,
the quantity $\xiskk$ can be determined by using
eqs.~(\ref{e8}), (\ref{e9}) and (\ref{e10}) up to a two-fold
ambiguity. This ambiguity can be resolved in principle,
if one takes into account the life-time splitting of the neutral
$B_s$-meson system which has been neglected in
eqs.~(\ref{e8})-(\ref{e10}) (for a discussion of this point see e.g.\
ref.~\cite{adk}).
Inserting $\xiskk$ extracted this way into the expression (\ref{e15}),
the quantity $z$ appearing in Fig.~1 can be determined provided the
CKM-angle $\gamma$ is known, for example, by applying the approach
proposed by Gronau and Wyler \cite{gw}. In contrast to the method
shown in Fig.~2, the approach using eq.~(\ref{e15}) to determine
$z$ suffers from $SU(3)$-breaking corrections that are related to
the spectator $s$-quark of the decaying $B_s$-meson~\cite{ghlrsu3}.
A reliable theoretical treatment of these corrections is
unfortunately not possible at present.

Let us note that one can extract in principle both $\gamma$ and the
amplitude $z$ simultaneously by combining Fig.~2 with eq.~(\ref{e15}).
This approach requires both time-independent measurements of the
branching ratios BR$(\bdpimkp)$, BR$(\bdbpipkm)$,
BR$(\bppipko)=\mbox{BR}(\bmpimkob)$ and a time-dependent measurement of
the CP asymmetry $\acp(t)$ of the decay $\bskk$ that has been defined
by eq.~(\ref{e8}). From the experimental point of view this
simultaneous approach seems, however, to be quite difficult.

Using the amplitude $z$ determined by applying either the approach
shown in Fig.~2 or the time-dependent CP asymmetry of the decay
$\bskk$, the $\bbtosb$ electroweak penguin amplitude
$(c_u-c_d)\pewp$ can be extracted with the help of Fig.~1. If one
follows Fig.~2 to determine $z$ and {\it defines} $(c_u-c_d)\pewp$ as the
electroweak penguin contribution to the decays $B^\pm\to\pi^0K^\pm$,
$SU(3)$-breaking does not affect the determination of this amplitude,
since we have only to deal with $B_{u,d}$ decays into $\pi K$ final
states. Consequently, besides the corrections related to the neglect
of the $C'$ and $\pewpc$ topologies (see eq.~(\ref{e2})), there are only
isospin-breaking corrections present in this approach.

Since electroweak penguins are dominated to a good approximation
by internal top-quark exchanges -- in contrast to the situation
concerning QCD penguins~\cite{bf} -- the $\bbtodb$ electroweak penguin
amplitude $(c_u-c_d)\pew$ is related in the limit of an exact $SU(3)$
flavour symmetry of strong interactions to the corresponding $\bbtosb$
amplitude through the relation
\beq\lab{e16}
(c_u-c_d)\pew=-\lambda R_te^{-i\beta}(c_u-c_d)\pewp.
\eeq
Here, $\lambda$ is the usual Wolfenstein parameter (in contrast to
the parameter $\bar\lambda$ in
eq.~(\ref{e2})) and $R_t$ represents the side of the unitarity triangle
that is related to $B^0_d$--$\bar B^0_d$--mixing. It is given by the
CKM-combination
\beq\lab{e17}
R_t\equiv\frac{1}{\lambda}\frac{|V_{td}|}{|V_{cb}|}.
\eeq
{}From present experimental data, we expect $R_t$ being of $\order(1)$
\cite{blo}.

The $\bbtodb$ electroweak penguin amplitude being
$\order(\bar\lambda^2)$ is mainly interesting
in connection with a clean determination of the CKM-angle $\alpha$
by using isospin relations among $B\to\pi\pi$ decays~\cite{gl}.
As we have pointed out already, although electroweak penguins
are expected to lead to small effects in this case
\cite{dhewp2,ghlrewp} it is an interesting and important question to
control the corresponding corrections quantitatively.

In Fig.~3 we have drawn the $B(\bar B)\to\pi\pi$ isospin triangles
in a way which differs from the one given in ref.~\cite{ghlrewp}
in order to illustrate the electroweak penguin corrections more
clearly. In particular we have rotated the $A(\bar B\to\pi\pi)$
amplitudes by the phase factor $e^{-2i\beta}$, which allows to
rotate $\pcewb$ back to $\pcew$:
\beq\lab{e18}
e^{-2i\beta}\pcewb=\pcew.
\eeq
This equation expresses the fact that the electroweak penguins are
dominated by internal top-quark exchanges. The angle $\phi$ appearing
in Fig.~3 fixing the relative orientation of the $B\to\pi\pi$
and $\bar B\to\pi\pi$ triangles is measured directly by the
mixing-induced CP asymmetry of the decay $B_d\to\pi^+\pi^-$ given by
\beq\lab{e19}
\acpmi(B_d\to\pi^+\pi^-)=-\frac{2|A(\bar B^0_d\to\pi^+\pi^-)|
|A(B^0_d\to\pi^+\pi^-)|}{|A(\bar B^0_d\to\pi^+\pi^-)|^2+
|A(B^0_d\to\pi^+\pi^-)|^2}\sin\phi
\eeq
which enters a formula for the corresponding time-dependent CP asymmetry
in an analogous way as in eq.~(\ref{e8}). Note that we would have
$\acpmi(B_d\to\pi^+\pi^-)=-\sin2\alpha$ and, thus, $\phi=2\alpha$, if
we neglected the penguin contributions to the decay $B_d\to\pi^+\pi^-$
completely.

Consequently, measuring both the $B(\bar B)\to\pi\pi$ rates
and the asymmetry $\acpmi(B_d\to\pi^+\pi^-)$, the solid and dashed
triangles shown in Fig.~3 can be constructed and the angle
$\tilde\alpha$ can be determined. This approach differs from the
original proposal of Gronau and London \cite{gl} (see also
ref.~\cite{ghlrewp}). Applying elementary trigonometry,
we find that the CKM-angle $\alpha$ is related to $\tilde\alpha$ through
\beq\lab{e20}
\alpha=\tilde\alpha+\Delta\alpha,
\eeq
where $\Delta\alpha$ is given by
\beq\lab{e21}
\Delta\alpha=r\sin\alpha\cos(\rho-\alpha)+\order(r^2)
\eeq
with
\beq\lab{e22}
r\equiv\frac{|(c_u-c_d)(\pew+\pewc)|}{|T+C|}.
\eeq
The phase $\rho$ is defined by
\beq\lab{e23}
(c_u-c_d)(\pew+\pewc)\equiv e^{i\rho}r(T+C).
\eeq
While it has been shown in ref.~\cite{ghlrewp} that
$\Delta\alpha=\order(r)$, we have calculated this correction
{\it quantitatively} in eq.~(\ref{e21}).

Taking into account that $\pewc/\pew,C/T=\order(\bar\lambda)$
\cite{ghlrewp,ghlrsu3} and employing both eq.~(\ref{e16}) and the
$SU(3)$-relation $T'=r_u T$ with $r_u\equiv
V_{us}/V_{ud}\approx\lambda$ \cite{grl}-\cite{ghlrsu3},
we find
\bea
r&\approx&\lambda r_u R_t\frac{|(c_u-c_d)\pewp|}{|T'|}\lab{e24}\\
\rho&\approx&\rho'-\beta+\pi,\lab{e25}
\eea
where $\rho'$ is a phase that is related to the $\bbtosb$ electroweak
penguin amplitude and that is defined in analogy to eq.~(\ref{e23})
through
\beq\lab{e26}
(c_u-c_d)\pewp\equiv e^{i\rho'}\frac{|(c_u-c_d)\pewp|}{|T'|}T'=
e^{i\rho'}\frac{|(c_u-c_d)\pewp|}{|z|}z.
\eeq
Strategies for the determination of the quantity $z$ have been discussed
above (see Fig.~2 or eq.~(\ref{e15})).

Consequently, inserting (\ref{e24}) and (\ref{e25}) into (\ref{e21})
and using the relation $\gamma=\pi-\alpha-\beta$, we obtain
\beq\lab{e27}
\Delta\alpha\approx\lambda r_u R_t\frac{|(c_u-c_d)\pewp|}{|T'|}
\sin\tilde\alpha\cos(\rho'+\gamma).
\eeq
Note that replacing $\sin\alpha$ appearing in eq.~(\ref{e21})
by $\sin\tilde\alpha$ leads to corrections of $\order(r^2)$ which
have been neglected in eq.~(\ref{e27}). The nice feature of this equation
is related to the fact that it includes only quantities that can
be determined by using Figs.~1--3 ($\gamma$ is one
of our inputs). Therefore, using this expression we are in a position
to estimate the electroweak penguin contribution to the value of
$\tilde\alpha$ in a quantitative way and consequently we
can extract the CKM-angle $\alpha$ with the help of eq.~(\ref{e20}).

At this point a discussion of $SU(3)$-breaking effects seems to be
in order. Whereas factorizable $SU(3)$-breaking affecting the relation
between $T'$ and $T$ can be included straightforwardly by setting
$r_u=\lambda f_K/f_\pi$~\cite{grl}-\cite{ghlrsu3},
such corrections on eq.~(\ref{e16}) are
more difficult to estimate as they involve not only meson decay
constants but also hadronic form factors.
Approximately, factorizable $SU(3)$-breaking can be taken into account
in this equation by multiplying its r.h.s.\ by the factor
$F_{B\pi}(0;0^+)/F_{BK}(0;0^+)$, where $F_{B\pi}(0;0^+)$ and
$F_{BK}(0;0^+)$ are form factors
parametrizing the hadronic quark-current matrix elements $\langle\pi^+
|(\bar bd)_{\VmA}|B^+\rangle$ and $\langle K^+|(\bar bs)_{\VmA}|B^+
\rangle$, respectively \cite{bgr}. Combining these considerations,
we obtain the following expression  for $\Delta\alpha$:
\beq\lab{e28}
\Delta\alpha\approx\left[\frac{f_K}{f_\pi}\frac{F_{B\pi}(0;0^+)}
{F_{BK}(0;0^+)}\right]\left[\lambda^2 R_t\frac{|(c_u-c_d)\pewp|}
{|T'|}\sin\tilde\alpha\cos(\rho'+\gamma)\right],
\eeq
which includes factorizable $SU(3)$-breaking in an approximate way.
At present there is no reliable theoretical technique available to
calculate non-factorizable $SU(3)$-breaking corrections to this
expression.

\newpage

Let us summarize briefly the main results of this letter:
\begin{itemize}
\item Using the CKM-phase $\gamma$ as an input and making some
reasonable approximations, we have shown that the
$\bbtosb$ electroweak penguin amplitude $(c_u-c_d)\pewp$ can be
straightforwardly determined.
\item To this end, one has to measure the three branching
ratios BR$(\bppiokp)$, BR$(\bmpiokm)$, BR$(\bppipko)=
\mbox{BR}(\bmpimkob)\propto|P'|^2$ and has, moreover, to determine
the amplitude $z\equiv T'/|P'|$.
\item We have presented two different strategies for extracting
the quantity $z$:
\begin{itemize}
\item A geometrical construction using the branching ratios
BR$(\bdpimkp)$ and BR$(\bdbpipkm)$.
\item A more formal method using the time-dependent CP asymmetry of
the mode $B_s\to K^+K^-$.
\end{itemize}
Whereas the latter approach suffers from $SU(3)$-breaking corrections
that are related to the spectator $s$-quark of the decaying $B_s$-meson,
there is no $SU(3)$-breaking present in the former one and in the
corresponding determination of $(c_u-c_d)\pewp$, if one defines this
amplitude as the electroweak penguin contribution to the decays
$B^\pm\to\pi^0K^\pm$.
\item Note that all branching ratios involved are expected to be of
$\order(10^{-5})$ and should be available in the foreseeable future.
A measurement of the time-evolution of the decay $B_s\to K^+K^-$
will, however, be rather difficult.
\item As electroweak penguins are dominated by internal top-quark
exchanges, we obtain a simple $SU(3)$-relation between the $\bbtosb$
and $\bbtodb$ electroweak penguin amplitudes.
\item Using this relation and the experimentally determined amplitude
$(c_u-c_d)\pewp$ we are in a position to estimate the
electroweak penguin contribution $\Delta\alpha$ to the angle
$\tilde\alpha=\alpha-\Delta\alpha$. This angle can be determined
following the approach of Gronau and London \cite{gl} by
measuring the branching ratios of the decays
$B(\bar B)\to\pi\pi$ and the CP asymmetry
$\acpmi(B_d\to\pi^+\pi^-)$. Therefore, the CKM-angle
$\alpha$ can be extracted and the electroweak penguin corrections
are -- at least in principle -- under control.
\end{itemize}
The possibility of determining $\pewp$ and $\pew$ experimentally as
suggested here opens the door for a quantitative study of electroweak
penguin effects in other $B$-decays. We will return to this in
a separate publication~\cite{bfewpapp}.

\vspace{1cm}

\section*{Figure Captions}
\begin{table}[ht]
\begin{tabular}{ll}
Fig.\ 1:&A geometrical strategy for determining the $\bbtosb$
electroweak pen-\\
&guin amplitude $(c_u-c_d)\pewp$.\\
&\\
Fig.\ 2:&The determination of the amplitude $z$ by using the modes
$B_d\to$\\
&$\pi^-K^+$, $\bdbpipkm$ and $\bppipko$ to fix $|P'|$.\\
&\\
Fig.\ 3:&The determination of the angle $\tilde\alpha$ by using
$B(\bar B)\to\pi\pi$ decays and\\
&its relation to the CKM-angle $\alpha$.\\
&\\
\end{tabular}
\end{table}

\newpage

\begin{figure}[p]
\vspace{15cm}
\caption{}\lab{f1}
\end{figure}

\begin{figure}[p]
\vspace{15cm}
\caption{}\lab{f2}
\end{figure}

\begin{figure}[p]
\vspace{15cm}
\caption{}\lab{f3}
\end{figure}

\end{document}